\documentstyle[aps, psfig]{revtex}

\begin{document}

\title{Phenomenological damping in trapped atomic Bose-Einstein condensates}
\author{S. \ Choi,  \ S. A. \ Morgan,  and K.\ Burnett}
\address{Clarendon Laboratory, 
Department of Physics, University of Oxford, Parks Road, 
\mbox{Oxford OX1 3PU, United Kingdom.}}

\vspace{6mm}
\maketitle
\begin{abstract}
The method of phenomenological damping developed by Pitaevskii for superfluidity near the $\lambda$ point is simulated numerically for the case of a dilute, alkali, inhomogeneous Bose-condensed gas near absolute zero. We study several features of this method in describing the damping of excitations in a Bose-Einstein condensate. In addition, we show that the method may be employed to obtain numerically accurate ground states for a variety of trap potentials.
\end{abstract}


Recent experiments with dilute, trapped Bose-Einstein condensates (BEC) have indicated the presence of damping of collective excitations\cite{JinMatEns97,MewAndvan96}. The most commonly used theoretical tool so far for describing BEC, 
the Gross-Pitaevskii equation (GPE)\cite{GPE}, has been shown to give accurate predictions for the frequencies of elementary excitations\cite{EdwRupBur96,JinEnsMat96,Str96}, but does not contain any mechanism for the damping of these excitations. 
Theories that go beyond the GPE are now subjects of intense study\cite{Gri96,ProBurSto98}. One of the theories which incorporates dissipation in a quantum fluid is Pitaevskii's phenomenological theory of superfluidity near  the $\lambda$ point\cite{Pit59}, and we have used it to investigate  damping in a dilute, trapped BEC numerically. This phenomenological method of including damping should, as argued by Pitaevskii, be valid when the system is close to thermal equilibrium.  We shall begin by deriving the required modification of GPE. We intend to demonstrate  in this paper two distinct uses of the equation: its use in describing damping of oscillations in trapped atomic BEC, and its use as a computational method to generate accurate eigenstates of the standard GPE. 

The standard GPE describes the motion of the mean field, $\psi$, which is taken to be the ensemble average of the boson field operator $\hat{\Psi}$. We write the GPE in the form
\begin{equation}
i \hbar \frac{\partial \psi}{\partial t}  =  -\frac{\hbar^2}{2m}\nabla^{2}_{\bf r}\psi +  V_{\rm trap}({\bf r})\psi + C_{n} |\psi|^2 \psi,
\end{equation}
with $C_{n}$ defined by
\begin{equation}
C_{n} = NU_{0} = \frac{4\pi\hbar^{2} N a}{m},
\end{equation}
where $N$ is the number of particles, $a$ the interatomic $s$-wave scattering length and $m$ the mass of a single particle.
The mean field $\psi$ is customarily taken to be the condensate wave function, although it is also consistent to define $\psi$ to be the condensate  ground state plus any {\em coherent} excitation. 

From inspection, the GPE contains no term which can describe damping. The first two terms are simply the energy of a single particle in the trap while the third  describes the non-dissipative effect of other particles in the system.
Attempts to improve on the GPE include taking into account the correlation effects of excitations\cite{Gri96} and the introduction of the many-body $T$-matrix\cite{ProBurSto98}.  These theories so far have not produced any explicit results for damping in the inhomogeneous case, although recent related work by Giorgini has shown the presence of Landau and Beliaev damping\cite{Gio97}.   

It is, however, possible to include damping phenomenologically, in a manner similar to that introduced by Pitaevskii\cite{Pit59}. The argument which we present below to obtain our expression for phenomenological damping follows this work closely. The procedure described is, in fact, a widely used one in the more general problems of dynamical critical phenomena\cite{LLPK}.

First of all, we note that any damping process eventually leads to an equilibrium state. For the condensate at $T=0$ such an equilibrium is described by the equation 
\begin{equation}
-\frac{\hbar^2}{2m}\nabla^{2}_{\bf r}\psi +  V_{\rm trap}({\bf r})\psi + C_{n} |\psi|^2 \psi  - \mu \psi = 0.   \label{TIGPE}
\end{equation}
This expression may be derived from minimizing an appropriate energy functional or equivalently by finding the eigensolution to the time dependent GPE.   
Because the process we want to describe is a relaxation process, in our equation of motion for the condensate $\psi$, 
\begin{equation}
i \hbar \frac{\partial \psi}{\partial t}  =  \hat{{\cal L}} \psi,
\end{equation}
the operator $\hat{{\cal L}}$ cannot be Hermitian. The anti-Hermitian part of $\hat{{\cal L}}$ is associated with the processes by which equilibrium is approached.  This anti-Hermitian part must vanish at equilibrium, that is, when Eq.~(\ref{TIGPE}) is satisfied.  For the cases where the deviation from equilibrium is small, one may therefore write the anti-Hermitian part in the form
\begin{equation}
i \Lambda \left (\frac{\hbar^2}{2m}\nabla^{2}_{\bf r}\psi +  V_{\rm trap}({\bf r})\psi + C_{n} |\psi|^2 \psi  - \mu \psi \right ),
\end{equation}
where $\Lambda$ is a dimensionless factor which is inversely proportional to the relaxation time. The Hermitian part is given by the fact that for $\Lambda=0$, the theory must reduce to the conventional GPE at $T=0$; the final equation including relaxation to equilibrium is, then, given by 
\begin{equation}
i \hbar \frac{\partial \psi}{\partial t} = (1+ i \Lambda) \left \{-\frac{\hbar^2}{2m}\nabla^{2}_{\bf r} +  V_{\rm trap}({\bf r}) + C_{n} |\psi|^2  - \mu \right \}\psi,  \label{pit}
\end{equation}
where $\Lambda < 0$ for damping. 

Equation (\ref{pit}) is, in fact, a rather general equation of motion which describes evolution towards equilibrium. Damping of elementary excitations may be described in the frame work of Eq. (\ref{pit}) by considering our  mean field $\psi$ to be broken into two parts: the condensate ground state $\psi_{g}$ and a {\em coherent} excitation, $\delta$,
\begin{equation}
\psi = e^{-i \mu t} ( \psi_{g} + \delta ).
\end{equation}
Evolution in Eq. (\ref{pit}) results in the damping of $\delta$ towards zero. The exact value for the damping parameter $\Lambda$ is expected to depend on a variety of factors. One would, for instance, expect it
to be temperature dependent, since the thermal component present in the trap is clearly going to be one of the main sources of damping. 
In the original case of absorption of sound in superfluid Helium considered by Pitaevskii, $\Lambda$ could be expressed in terms of the coefficients of second viscosity used by Khalatnikov\cite{Kha52}. A closely related result (which actually derives the Landau-Khalatnikov two-fluid model for the homogeneous Bose-condensed gas) has been calculated by Kirkpatrick and Dorfman using the Chapman-Enskog procedure\cite{KirDor83}. They also explicitly derived various  transport coefficients such as the viscosity for their homogeneous case. At present, a complete analysis for the inhomogeneous case is not yet available.

Some rough initial estimate for the damping coefficient $\Lambda$ may, however, be made by noting that, physically, $\Lambda$ represents the rate at which the excited components turn into the condensate. This may be approximated by the transition probability $W^{+}(N)$ that has been estimated in a recent work by Gardiner {\it et al.}\cite{GarZolBal97} using the quantum kinetic theory.  $W^{+}(N)$ gives the rate at which the thermal particles above the condensate band enter the condensate due to collisions. Although the description of the process of evaporative cooling was their primary goal, their main assumption in formulating $W^{+}(N)$ was that the condensate does not readily act back on the thermal component to change its temperature. This assumption is expected to be valid when one considers quasiequilibrium situations.   Using the reported values from the MIT experiment\cite{MewAndvan96} as an example, and the expression 
\begin{equation}
W^{+}(N) = \frac{4m(akT)^{2}}{\pi \hbar^{3}}e^{2 \mu/kT} \left \{ \frac{\mu_{N}}{kT}K_{1}\left (\frac{\mu_{N}}{kT} \right ) \right \},
\end{equation}
where $K_{1}(z)$ is a modified Bessel function,  we obtain $\Lambda \sim W^{+}(N)/\omega \approx -0.03$, for $T \approx T_{c}/10$ and $a \approx 3.45 {\rm nm}$. Here, $\omega$ was taken to be the trap frequency relevant to the oscillatory motion, that is $\omega = 2 \pi \times 19 {\rm Hz}$. The corresponding damping time is then approximately 200~ms, which is of the same order of magnitude as the experimental damping time of 250~ms\cite{MewAndvan96}. 

Numerical simulation of Eq.~(\ref{pit}) is done using a 4th order Runge-Kutta algorithm. Although the simulations in this paper are performed using a one dimensional (1D) Runge-Kutta integrator, the general arguments that follow do not depend on dimensionality, and  it is straight forward to apply the ideas of the present paper to a 3D simulation.  
We simulate damping of elementary excitations by first populating a mode of elementary excitation on a condensate and then time-evolving this condensate plus excitation using Eq. (\ref{pit}). In practice, such a state is generated by, for instance, applying time dependent perturbations to the confining potential initially and then observing the subsequent behaviour of the condensate after turning off the perturbation. Exactly how the excitations damp using this formalism can be analysed quantitatively using the method of quasiparticle projection\cite{MorChoiBur97}.  This projection method is based on the Bogoliubov transformation of fluctuations about the condensate mean field:  one writes a general wave function $\psi$ as 
\begin{equation}
\psi = e^{-i \mu t} \left [(1+b_{g}) \psi_{g}({\bf r}) + \sum_{i} \left \{ u_{i}({\bf r}) b_{i} + v^{*}_{i}({\bf r}) b_{i}^{*} \right \} \right ]  \label{qp}
\end{equation}
where the index $i$ denotes quasiparticle energy levels, and the functions $u_{i}({\bf r})$ and $v_{i}({\bf r})$ obey a set of coupled equations usually known as the Bogoliubov-de Gennes equations which diagonalise a quadratic Hamiltonian\cite{Fet72}.   The sum over $i$ on the right hand side of Eq. (\ref{qp}) represents the fluctuation about the condensate ground state $\psi_{g}$ which is itself weighted by a factor $(1 + b_{g})$. The population of quasiparticles in level $i$ is then given by the quantity $b_{i}^{*}b_{i} = |b_{i}|^{2}$ where the coefficients $b_{i}$ are taken to be the mean values of the quasiparticle annihilation operators.  By using the well known orthogonality relations of $u_{i}({\bf r})$ and $v_{i}({\bf r})$, one may extract the expansion coefficients $b_{g}$ and $b_{i}$ uniquely at any point in time, and hence deduce the evolution of the quasiparticle population in each mode\cite{MorChoiBur97}.  
 
In order to simulate the experimental results as closely as possible, we initially populate the breathing mode, the second lowest excitation, by an arbitrary amount and then plot the evolution of this population, $|b_{2}|^{2}$, over time.  Figure 1 shows the population of mode 2 over time as well as the population in the condensate mode $|b_{g}|^{2}$ for various values of $\Lambda$.    An initial population of 0.25 means $|b_{2}|^2 = 0.25$. The corresponding ground state population is shown in the lower part of the plot.  We see that for smaller values of $\Lambda$, i.e. when the excited component does not damp out quickly over time, there is a greater interplay between the ground and the excited state populations due to nonlinear mixing. 

The width of the condensate is indeed what is actually measured in the experiment, and therefore this is plotted as a function of time in Fig. 2. The general shape reported in Ref. \cite{JinMatEns97} of the form $A \exp (-\Lambda t) \sin(2\pi \nu t + \phi) + B$ is clearly reflected in this figure. The solid curve gives what is expected for the MIT experiment\cite{MewAndvan96} (i.e. $\Lambda \approx -0.03$).  As the simulation was done in 1D, however, we do not intend to claim any quantitative analogy. Also, in real experiments, several quasiparticle modes, rather than just the breathing mode are simultaneously excited. It is clear, however, that this description of damping should be satisfactory in cases where an exact formalism is not needed: for instance, when $\Lambda$ may be treated as a parameter to be obtained from an experiment. 

One of the ways to characterise different descriptions of damping is to see its dependence on energy. For instance, in the homogeneous gas, it is well known that the damping rate increases as $k^2$ when the relaxation process is due to thermal conduction and viscosity\cite{LLPK}. We note that the Pitaevskii prescription gives a damping rate which has a linear dependence on the mode energy, with the high energy components being damped out more rapidly. This aspect may be seen simply by inspection of the equation of motion Eq.~(\ref{pit}), and it can also be confirmed by explicitly calculating the damping rate of the quasiparticle populations, $|b_{i}|^{2}$, for various values of $i$ by linearisation of Eq.~(\ref{pit}).  

We now look at the second possible use of Eq.~(\ref{pit}). We note that finding a reliable solution to the time-independent GPE is an essential first step in many numerical simulations involving a trapped BEC.  The Pitaevskii damping description may, in fact, be used to this end as an effective numerical tool to find the eigenstate. This is possible because the equilibrium state Eq.~(\ref{TIGPE}) is itself a solution to Eq.~(\ref{pit}).  To find the eigenstate, one starts by initially choosing an arbitrary function (preferably close to the desired solution for faster convergence, e.g. for a condensate in a trap, a Gaussian or a Thomas-Fermi solution) and run the routine with pre-determined values of $\mu$ and $\Lambda$.  It should be noted here that, since the equation is non-hermitian, the evolution does not conserve the norm of $\psi$. It is therefore necessary to proceed by first running the routine with an estimate of $C_{n}$ for the chosen $\mu$ until convergence, and then later adjusting $C_{n}$ from the renormalisation of $\psi$,
\begin{equation}
\int |\psi({\bf r})|^2 d{\bf r}= 1.
\end{equation} 
The correct $C_{n}$ for the chosen $\mu$ is then given by the initial estimate divided by the final normalisation constant.  

We found that the time it takes to find a solution in this way is
very much shorter than other methods such as the adiabatic turn on of the non-linearity  $C_{n}$. As a rough indication, evolution time of $t = 10/\omega$ was required for growing an eigenstate with $C_{n} = 100$ and $\Lambda = -2$ when using the damping method while it took us $t = 500/\omega$ using the adiabatic method; in real time on a Sun SPARC station 10, these running times corresponded to roughly 10 minutes and 3 hours respectively.  The damping method is able to find a solution for as large a $C_{n}$ as desired, provided there are a sufficient number of grid points to accommodate the larger condensate. An exact eigensolution to GPE has a flat phase profile; the damping simulation with $\Lambda = -2$, $C_{n} = 50$ running for $t = 15/\omega$ typically gave a maximum change in phase across the condensate of the order of $10^{-7}$ radians, indicating the high accuracy of this method. 

Other advantages of this approach include its flexibility in handling traps of any arbitrary shape, and the fact that it is not necessary  to expand the wave function in a set of complete, orthogonal basis states which can be a computationally expensive operation. 

The value of $\Lambda$ cannot, however, be increased indefinitely for faster convergence, as this results in numerical instability. 
It was found that we get numerical instability when $\Lambda  \/ t_{\rm step} > 0.03$  where $t_{\rm step}$ is the size of the incremental time step in numerical integration. The onset of numerical instability is independent of the total length of simulation and the size of the non-linearity constant $C_{n}$. One may understand this from the fact that the damping per time step is roughly given by $e^{-\Lambda t_{\rm step}}$ so that $e^{-0.03} \approx 0.97$ implies a reduction in the amplitude by about 3 \% per numerical time step.  Since the typical step size used is of the order of $10^{-3}$ harmonic oscillator units, this, in fact, represents a complete decay on a time scale of order a period. 
 
We also point out that the vortex eigenstate in a trap may be obtained numerically by using this damping prescription; one needs only to start with the first excited harmonic oscillator state which has an odd parity. One of the corollaries of the damping method is that although parity is conserved the number of zero crossings is not; if we started with, say, the 3rd excited harmonic oscillator state, the resulting solution is still the state with one zero crossing. The phase profile of such a condensate displays a clear $\pi$ phase jump at the origin, while being flat everywhere else. 

In summary, we found that the phenomenological damping description does indeed give results which are compatible with those of the experiments. In a more general context, it provides an efficient numerical tool for finding the eigenstate of the time independent GPE.  

\acknowledgments
SC would like to acknowlege the support of UK CVCP and the Wingate Foundation;  
SM and KB acknowledge the support from the UK EPSRC. KB also acknowledges support from the European Union under the TMR Network Programme.

\vspace{10mm} 

\begin{figure}[t] \begin{center}
\centerline{\psfig{height=7cm,file=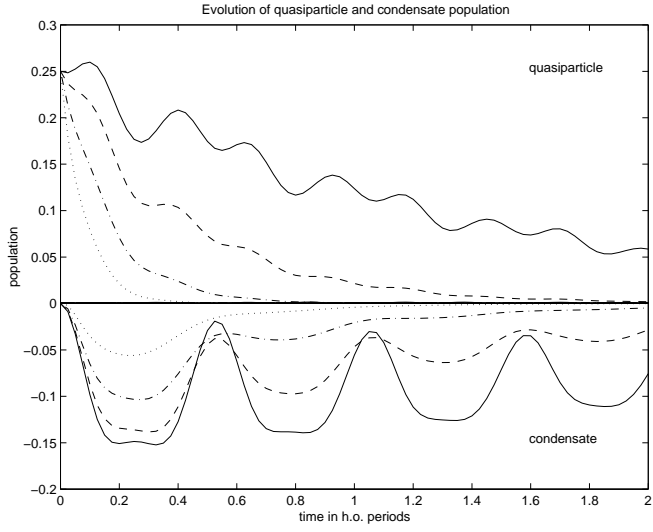}}
\end{center}
\caption{\protect \footnotesize    Population of the breathing mode, $|b_{2}|^{2}$,  over time  for various values of $\Lambda$.   Solid, dashed,  dot-dashed, and dotted  lines indicate $\Lambda$ of $-0.03$, $-0.1$, $-0.25$, and $-0.5$ respectively.  The lower half of the plot shows the corresponding variation in  $|b_{g}|^{2}$.}
\end{figure}

\begin{figure}[t] \begin{center}
\centerline{\psfig{height=7cm,file=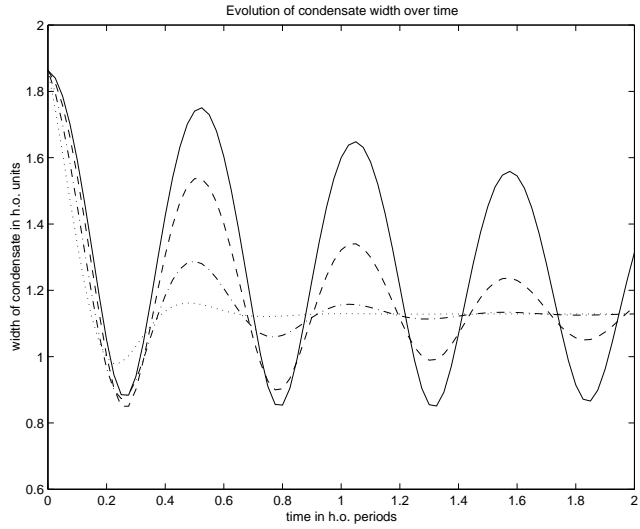}}
\end{center}
\caption{\protect \footnotesize  Variation of the width of the trapped condensate over time. The width is measured by the standard deviation of $x$, the position, weighted by the corresponding condensate density.  Solid, dashed,  dot-dashed and dotted lines indicate $\Lambda$ of $-0.03$, $-0.1$, $-0.25$, and $-0.5$ respectively, as in the previous figure. }
\end{figure}


\vspace{0mm}

\begin{references}

\bibitem{JinMatEns97}  D. S. Jin, M. R. Matthews,  J. R. Enscher, C. E. Wieman,  and E. A. Cornell, Phys. Rev. Lett. {\bf 78}, 764 (1997)

\bibitem{MewAndvan96} M.-O. Mewes, M. R. Andrews, N. J. van Druten, D. M. Kurn,  D. S. Durfee, C. G. Townsend, and W. Ketterle Phys. Rev. Lett. {\bf 77}, 988 (1996)

\bibitem{GPE}V. L. Ginzburg and L. P. Pitaevskii, Zh. Eksp. Teor. Fiz. {\bf 34}, 1240 (1958) [Sov. Phys. JETP {\bf 7}, 858 (1958)]; E. P. Gross, J. Math. Phys. {\bf 4}, 195 (1963)

\bibitem{EdwRupBur96} M. Edwards, P. A. Ruprecht, K. Burnett, R. J. Dodd, and C. W. Clark  Phys. Rev. Lett {\bf 77}, 1671 (1996)

\bibitem{JinEnsMat96}  D. S. Jin, J. R. Enscher, M. R. Matthews, C. E. Wieman, and E. A. Cornell, Phys. Rev. Lett. {\bf 77}, 420 (1996)

\bibitem{Str96} S. Stringari, Phys. Rev. Lett {\bf 77}, 2360 (1996)

\bibitem{Gri96} A. Griffin Phys. Rev. B {\bf 53}, 9341 (1996)

\bibitem{ProBurSto98} N. P. Proukakis, K. Burnett, and H. T. C. Stoof, To appear in Phys. Rev. A (1998)

\bibitem{Pit59}  L. P. Pitaevskii, Sov. Phys. J.E.T.P.  {\bf 35}, 282 (1959)

\bibitem{Gio97} S. Giorgini, To appear in Phys. Rev. A (1998), e-print cond-mat/9709259

\bibitem{LLPK}  E. M. Lifshitz and L. P. Pitaevskii,  Physical Kinetics, Landau and Lifshitz Course of Theoretical Physics Vol. 10 (Pergamon Press, Oxford 1981)

\bibitem{Kha52}  I. M. Khalatnikov J. Exptl. Theoret. Phys. (U.S.S.R.)  {\bf 23}, 8 (1952)

\bibitem{KirDor83}  T. R. Kirkpatrick and J. R. Dorfman, Phys. Rev. A  {\bf 28}, 2576 (1983)

\bibitem{GarZolBal97} C. W. Gardiner, P. Zoller, R. J. Ballagh, and M. J. Davis, Phys. Rev. Lett.  {\bf 79}, 1793 (1997)

\bibitem{MorChoiBur97} S. A. Morgan, S. Choi, K. Burnett, and M. Edwards submitted to Phys. Rev. A

\bibitem{Fet72} A. L. Fetter, Ann. Phys. {\bf 70}, 67 (1972)

\end{references}
\end{document}